\def\nk{n_{\rm b}}

\def\Pb{P_{\rm b}}

\def\rfr#1{Equation\,(\ref{#1})}
\def\rfrs#1#2{Equations\,(\ref{#1})-(\ref{#2})}
\def\Rfr#1{Equation\,(\ref{#1})}
\def\Rfrs#1#2{Equations\,(\ref{#1})-(\ref{#2})}
\def\derp#1#2{\rp{\partial{#1}}{\partial{#2}}}
\def\dert#1#2{\frac{{{\textrm{d}}}{#1}}{{{\textrm{d}}}{#2}}}

\def\virg#1{``#1"}

\def\eqi{\begin{equation}}
\def\eqf{\end{equation}}
\def\eqia{\begin{eqnarray}}
\def\eqfa{\end{eqnarray}}

\def\rp#1#2{{#1\over#2}}
\def\lb#1{\label{#1}}

\def\bds#1{\boldsymbol{#1}}

\def\ton#1{\left(#1\right)}
\def\qua#1{\left[#1\right]}
\def\grf#1{\left\{#1\right\}}

\documentclass[onecolumn]{aastex}
\usepackage{morefloats}
\usepackage[title]{appendix}
\usepackage{hyperref}
\usepackage{booktabs}
\usepackage[table,xcdraw]{xcolor}
\usepackage{multirow}
\usepackage{rotating,tabularx}
\usepackage{float}
\usepackage{enumerate}
\usepackage{rotating}
\usepackage[polutonikogreek,english]{babel}
\usepackage{amsmath,starfont,textgreek,w-greek,wasysym}
\usepackage{amsthm}
\usepackage{amscd,lineno}
\usepackage{amssymb,dsfont}
\usepackage{graphicx,epsfig}
\usepackage{txfonts}
\usepackage{xr-hyper}

\RequirePackage{color}

\newcommand{\emaila}{lorenzo.iorio@libero.it}

\allowdisplaybreaks[1]

\begin{document}

\title{Revisiting the 2PN pericentre precession in view of possible future measurements of it}

\shortauthors{L. Iorio}

\author{Lorenzo Iorio\altaffilmark{1} }
\affil{Ministero dell'Istruzione, dell'Universit\`{a} e della Ricerca
(M.I.U.R.)-Istruzione
\\ Permanent address for correspondence: Viale Unit\`{a} di Italia 68, 70125, Bari (BA),
Italy}

\email{\emaila}

\begin{abstract}
At the second post-Newtonian (2PN) order, the secular pericentre precession $\dot\omega^\mathrm{2PN}$  of either a full two-body system made of well detached non-rotating monopole masses of comparable size and a restricted two-body system composed of a point particle orbiting a fixed central mass have been analytically computed so far with a variety of approaches.
We offer our contribution by analytically computing $\dot\omega^\mathrm{2PN}$ in a perturbative way with the method of variation of elliptical elements by explicitly calculating both the direct contribution due to the 2PN acceleration ${\bds A}^\mathrm{2PN}$, and also an indirect part arising from the self-interaction of the 1PN acceleration ${\bds A}^\mathrm{1PN}$ in the orbital average accounting for the instantaneous shifts induced by ${\bds A}^\mathrm{1PN}$ itself. Explicit formulas are straightforwardly obtained for both the point particle and full two-body cases without recurring to simplifying assumptions on the eccentricity $e$. Two different numerical integrations of the equations of motion confirm our analytical results for both the direct and indirect precessions. The values of the resulting effects for Mercury and some binary pulsars are confronted with the present-day level of experimental accuracies in measuring/constraining their pericentre precessions. The supermassive binary black hole in the BL Lac object OJ 287 is considered as well. A comparison with some of the results appeared in the literature is made.
\end{abstract}

keywords{
general relativity and gravitation;  celestial mechanics; experimental studies of gravity; ephemerides
}
\section{Introduction}
The problem of calculating at the second post-Newtonian (2PN) order of general relativity \citep{2016Univ....2...23D} the secular\footnote{For the sake of simplicity, we will omit the brackets denoting the average over one orbital revolution  here and throughout the paper.} precession $\dot\omega^\mathrm{2PN}$ of pericentre $\omega$ of a full two-body system made of a pair of detached, non-rotating masses of comparable sizes and of a restricted two-body system characterized by a test particle orbiting its massive primary has been analytically tackled several times so far with a variety of calculational approaches
\citep{1976PThPh..56..324H,1987CRASB.305..839D,1988NCimB.101..127D,Nippo89,1993PhLA..174..196S,
1993PhLA..177..461S,1994ARep...38..104K,1995CQGra..12..983W,DONHAT1998328,PhysRevD.70.104011,
PhysRevD.71.024039,cinesi09,Eliseo011,2013MNRAS.428.1201B,Blanchet2014,
PhysRevD.91.024012,2017PhRvD..95f4003W,Marin018,Mak18,2019CQGra..36k5001T,2018MNRAS.480.3747W,2018PhRvL.120s1101W}.
In spite of their formal elegance, it is not always easy to extract from them quickly understandable formulas, ready to be read and used in practical calculations in view of possible confrontation with actual data from astronomical and astrophysical scenarios of potential experimental interest. Perhaps, it is so because, e.g., of continuous references nested one inside the other to various papers pointing to a host of intermediate parameterizations, often of purely theoretical relevance, that tend somehow to confuse a little bit at least some readers. Sometimes, they may wonder which numerical values of the parameters of the system under consideration out of those recorded in the literature have to be inserted in the equations.
For a recent discussion on some aspects of the approaches followed in the literature so far, see \citet{2019CQGra..36k5001T}; see also \citet{1994ApJ...427..951K} for a comparison of some of the parameterizations used in the literature to the 1PN level.

Our aim is revisiting the issue of analytically calculating the 2PN pericentre precession by straightforwardly computing it perturbatively with the widely known method of variation of the orbital elements \citep{Tiss1889,Plummer60,1961mcm..book.....B,Danby62,Sof89,1991ercm.book.....B,2000ssd..book.....M,2003ASSL..293.....B,2005ormo.book.....R,2011rcms.book.....K,2014grav.book.....P,SoffelHan19} in order to provide quickly understandable formulas, ready to be used in practical calculations in view of possible measurements in a not so far future, more likely in binary pulsars than in our Solar system, or to better model the dynamics of peculiar systems like, e.g., tight extrasolar planetary systems or the BL Lac object OJ 287 \citep{2018ApJ...866...11D,2019Univ....5..108D}. A similar strategy was adopted in \citet{1994ARep...38..104K}. Because the actual data analyses of astronomical and astrophysical systems are performed by using the harmonic coordinates of PN theory, we will adopt them in our calculation \citep[see the discussion in Sec.\,4 of][]{2019CQGra..36k5001T}. We will, first, deal with the point particle case (Section\,\ref{popaca}) by starting with the precession \textit{directly} induced by the 2PN acceleration ${\bds A}^\mathrm{2PN}$ entering the equations of motion (Section\,\ref{dir2PNpp}). Then, in Section\,\ref{indir2PNpp}, we will calculate the \textit{indirect} 2PN precession arising from the fact that, to the order $\mathcal{O}\ton{c^{-4}}$, where $c$ is the speed of light in vacuum, also the instantaneous shifts of the orbital elements occurring during an orbital revolution due to the 1PN acceleration ${\bds A}^\mathrm{1PN}$ itself should be taken into account in the averaging procedure of the 1PN effects. Instead, neglecting such changes gives rise to the usual, time-honored Einstein-like 1PN precession. In principle, also other general relativistic precessions  may be calculated, to the order $\mathcal{O}\ton{c^{-4}}$, from the mutual interaction of some 1PN accelerations induced by the bodies'  mass and spin moments \citep{1988CeMec..42...81S,1990CeMDA..47..205H,2014CQGra..31x5012P,2015CeMDA.123....1M,2018RSOS....580640F,2018CeMDA.130...40S} entering the equations of motion; they will not be treated here because of their smallness. For some of them, see \citet{2015IJMPD..2450067I}. Section\,\ref{numint} contains  numerical integrations of the equations of motion of some binary systems confirming our analytical result of Section\,\ref{dir2PNpp} for the direct effect, and of Section\,\ref{indir2PNpp} for the indirect one. It turns out that the direct 2PN perihelion precession of Mercury is smaller than the present-day observational accuracy in constraining any unmodeled perihelion precession of Mercury by about an order of magnitude or so. Currently, the 2PN equations of motion are not included in the dynamical models of the Solar system dynamics employed by the teams of astronomers producing the planetary ephemerides \citep{2014IPNPR.196C...1F,2018AstL...44..554P,2019NSTIM.109.....F}. In Section\,\ref{mAmB} we repeat our calculation for a full two-body system by calculating  both the direct (Section\,\ref{dir2PNmAmB}) and the indirect (Section\,\ref{indir2PNmAmB}) contributions to the 2PN pericentre precession in the same fashion as in Section\,\ref{popaca}. We compute them for the Hulse-Taylor binary pulsar PSR B1913+16 \citep{1975ApJ...195L..51H} and the double pulsar PSR J07373039A/B \citep{2003Natur.426..531B,2004Sci...303.1153L} by comparing the resulting predictions with the current experimental accuracy in determining their periastron precessions from timing measurements. While for PSR B1913+16 the overall 2PN periastron precession is already potentially measurable today, for PSR J07373039A/B the indirect contribution, which depends explicitly on the initial value of the orbital phase, may weaken or even cancel out the direct effect for certain values of the initial position of the pulsar along its orbit. On the other hand, for other initial positions the total 2PN periastron precession may be brought above the measurability threshold.
We look also at the supermassive binary black hole in OJ 287.
In Section\,\ref{parago}, we compare our calculation with those in \citet{1994ARep...38..104K} by disclosing an error in their results for the indirect effects. A comparison is made also with  the results by \citet{1988NCimB.101..127D}.
Section\,\ref{fine} summarizes our findings, and offers our conclusions.
\section{The point particle case}\lb{popaca}
\subsection{The direct pericentre precession due to the 2PN acceleration}\lb{dir2PNpp}
The 2PN acceleration experienced by a test particle orbiting  a fixed  body of mass $M$ at distance $r$, written in harmonic coordinates, is \citetext{\citealp[see, e.g.,][Eq.\,(2.3)]{2017PhRvD..95f4003W}; \citealp[Eq.\,(8.8.16),\,p.\,332]{SoffelHan19}}
\eqi
\bds{A}^\mathrm{2PN} = \rp{\mu^2}{c^4\,r^3}\qua{\ton{2\,{\mathrm{v}}_r^2-\rp{9\,\mu}{r}}\bds{\hat{r}} - 2\,{\mathrm{v}}_r\,{\mathbf{v}} }.\lb{A2PN}
\eqf
In \rfr{A2PN}, $\mu\doteq GM$
is the gravitational parameter of the primary, $G$ is the Newtonian gravitational constant,
and ${\mathrm{v}}_r \doteq {\mathbf{v}}\bds\cdot\bds{\hat{r}} $
is the radial velocity of the test particle, i.e. the projection of its velocity ${\mathbf{v}}$ onto the versor $\bds{\hat{r}}$ of its position vector $\bds r$ with respect to the primary.
\Rfr{A2PN} can be obtained from the point particle limit of the 2PN equation of relative motion of a full two-body system  treated in Section\,\ref{dir2PNmAmB}. \Rfr{A2PN} can also be inferred  from the equation of motion of Equation\,(4.4.18) of \citet[p.\,152]{1991ercm.book.....B} or Equation\,(1.5c) of \citet[p.\,133]{1988NCimB.101..127D} for the body 1 assumed as test particle orbiting the body 2 taken as its primary, i.e. for $M_2\rightarrow M,\,{\mathbf{v}}_2\rightarrow 0,\,M_1\rightarrow 0,\,{\mathbf{v}}_1\rightarrow {\mathbf{v}}$.

Let us analytically work out the direct long-term, i.e. averaged one orbital period $\Pb$, 2PN precession of pericentre induced solely by \rfr{A2PN} by means of the Gauss equations \citep[e.g.][]{2011rcms.book.....K,2014grav.book.....P,SoffelHan19}, valid for any additional acceleration $\bds A$ with respect to the Newtonian monopole $A_\mathrm{N}=-\mu/r^2$,
\begin{align}
\dert\Omega t \lb{gaus1}& = \rp{r\,A_\nu\,\sin u}{\nk\,a^2\,\sqrt{1-e^2}\,\sin I}, \\ \nonumber \\
\dert\omega t \lb{gaus2}& = \rp{\sqrt{1-e^2}}{\nk\,a\,e}\qua{-A_r\,\cos f + A_\tau\ton{1+\rp{r}{p}}\sin f}-\cos I\,\dert\Omega t,
\end{align}
where $a,\,e,\,I,\,\Omega,\,\omega,\,f$ are the semimajor axis, eccentricity, inclination, longitude of the ascending node, argument of pericentre, and true anomaly, respectively, $p\doteq a\ton{1-e^2}$ is the semilatus rectum, $u\doteq\omega+f$ is the argument of latitude, $\nk\doteq\sqrt{\mu/a^3}$ is the Keplerian mean motion, while $A_r,\,A_\tau,\,A_\nu$ are the radial, transverse and out-of-plane components of the extra-acceleration $\bds A$, respectively.
It is appropriate to remark that the Gauss equations are exact since the possible smallness of  $\bds A$ with respect to $A_\mathrm{N}$ is not assumed  in their derivation \citep[p.\,108]{SoffelHan19}. In a perturbative calculation, which is fully adequate for the 2PN acceleration of \rfr{A2PN} in most of the situations in which a conceivable future detection could be envisaged (our Solar system, exoplanets, binary pulsars), the right-hand sides of \rfrs{gaus1}{gaus2} have to be evaluated onto the Keplerian ellipse $r=p/\ton{1+e\,\cos f}$, assumed as unperturbed, reference trajectory, and averaged out over one orbital period $\Pb\doteq 2\uppi/\nk$ by means of \citep{1958SvA.....2..147E,1959ForPh...7S..55T,1961mcm..book.....B,1970CeMec...2..369R,1979AN....300..313M,1991ercm.book.....B,2014grav.book.....P}
\eqi
\dert{t}{f} = \rp{r^2}{\sqrt{\mu\,p}}\,\rp{1}{1-\rp{r^2}{\sqrt{\mu\,p}}\,\ton{\dert\omega t + \cos I\,\dert\Omega t}}\simeq \rp{r^2}{\sqrt{\mu\,p}}\,\qua{1+\rp{r^2}{\sqrt{\mu\,p}}\,\ton{\dert\omega t + \cos I\,\dert\Omega t}}.\lb{dtdf}
\eqf
In it, the derivatives of $\omega$ and $\Omega$ are given by \rfrs{gaus1}{gaus2}.
In order to keep only terms of order $\mathcal{O}\ton{c^{-4}}$ when \rfr{A2PN} is used in \rfrs{gaus1}{gaus2}, only the first term of \rfr{dtdf} has to be retained because of the presence of $A$ itself in it through $\mathrm{d}\Omega/\mathrm{d}t,\,\mathrm{d}\omega/\mathrm{d}t$. It is intended that, in the following, the right-hand-sides of \rfrs{gaus1}{dtdf} are evaluated onto the constant Keplerian ellipse; in order to avoid an excessively cumbersome notation, we avoid to append a subscript $\virg{\mathrm{K}}$ to the orbital elements entering them.

The radial, transverse, and out-of-plane components of \rfr{A2PN}, evaluated onto the reference Keplerian trajectory,  turn out to be
\begin{align}
A_r^\mathrm{2PN} \lb{Ar2PN}& = -\rp{9\,a^5\,\nk^6\,\ton{1+e\,\cos f}^4}{c^4\,\ton{1-e^2}^4}, \\ \nonumber \\
A_{\tau}^\mathrm{2PN} \lb{At2PN}& = -\rp{2\,e\,a^5\,\nk^6\,\ton{1+e\,\cos f}^4\,\sin f}{c^4\,\ton{1-e^2}^4}, \\ \nonumber \\
A_{\nu}^\mathrm{2PN} \lb{An2PN} & = 0.
\end{align}
By inserting \rfrs{Ar2PN}{An2PN} into \rfrs{gaus1}{gaus2} and averaging with the first term of \rfr{dtdf} yields, to order $\mathcal{O}\ton{c^{-4}}$, the direct 2PN pericentre precession
\eqi
\dot\omega^\mathrm{2PN}_\mathrm{dir} = \rp{\nk\,\mu^2\,\ton{28 - e^2}}{4\,c^4\,a^2\,\ton{1-e^2}^2},\lb{dir2PN}
\eqf
corresponding to a shift per orbit
\eqi
\Delta\omega^\mathrm{2PN}_\mathrm{dir} = \rp{\uppi\,\mu^2\,\ton{28 - e^2}}{2\,c^4\,a^2\,\ton{1-e^2}^2}.
\eqf
The analytical result of \rfr{dir2PN} will be numerically confirmed in Section\,\ref{numint} by numerically integrating the equations of motion.
\subsection{The indirect pericentre precession due to the 1PN acceleration}\lb{indir2PNpp}
\rfr{dir2PN}, although directly inferred from the 2PN acceleration of \rfr{A2PN}, does not exhaust the issue of calculating the full pericentre precession to the order $\mathcal{O}\ton{c^{-4}}$. Indeed, there are also other two contributions to it, which may be dubbed as \virg{indirect}, coming from the well known 1PN acceleration itself \citep[e.g.\,][p.\,332]{SoffelHan19}
\eqi
{\bds{A}}^\mathrm{1PN} = \rp{\mu}{c^2\,r^2}\,\qua{\ton{\rp{4\,\mu}{r} -\mathrm{v}^2 }\,\bds{\hat{r}} +4\,{\mathrm{v}}_r\,{\mathbf{v}} }\lb{A1PN}.
\eqf
Basically, they arise because during an orbital revolution of the test particle under the perturbing influence of $\bds A$ like \rfr{A1PN} all the orbital elements, in principle, undergo instantaneous variations changing their values from their fixed Keplerian ones referred to some reference epoch $t_0$. Moreover, when the integration over $f$ is performed in order to obtain the net change per orbit, the fact that $f$ is reckoned from a generally varying line of apsides because of $\bds A$ should be taken into account as well. Such features yield additional corrections of the order of $\mathcal{O}\ton{A^2}$ which, in the present case, are just of the order of $\mathcal{O}\ton{c^{-4}}$. We will implement such a strategy by following \citet{2015IJMPD..2450067I} in which the indirect effects of order $\mathcal{O}\ton{J_2\,c^{-2}}$, where $J_2$ is the primary's oblateness, were computed in agreement with  \citet{2014PhRvD..89d4043W,2015PhRvD..91b9902W}.

One of the aforementioned indirect contributions to the 2PN pericentre precession, marked conventionally with the superscript $\ton{\mathrm{I}}$ in the following, is obtained from the orbital average of \rfrs{gaus1}{gaus2}, calculated with \rfr{A1PN}, by means of the second and third terms of \rfr{dtdf} containing just \rfr{A1PN} itself  which, among other things, shifts slowly the apsidal line from which the true anomaly $f$ is counted.
By recalling that the radial, transverse, and out-of-plane components of \rfr{A1PN} are \citep[Eq.\,(8.8.5)-(8.8.6),\,p.\,330]{SoffelHan19}
\begin{align}
A^\mathrm{1PN}_r \lb{A1PNr} & = \rp{\mu^2\,\ton{1 + e\,\cos f}^2\,\ton{3 + e^2 + 2\,e\,\cos f - 2\,e^2\,\cos 2 f}}{c^2\,a^3\,\ton{1-e^2}^3},\\ \nonumber \\
A^\mathrm{1PN}_\tau \lb{A1PNt} & = \rp{4\,e\,\mu^2\,\ton{1 + e\,\cos f}^3\,\sin f}{c^2\,a^3\,\ton{1-e^2}^3}, \\ \nonumber \\
A^\mathrm{1PN}_\nu \lb{A1PNn} & = 0,
\end{align}
the resulting indirect precession $\dot\omega_\mathrm{indir}^\mathrm{2PN\,\ton{I}}$ of order $\mathcal{O}\ton{c^{-4}}$ turns out to be
\eqi
\dot\omega_\mathrm{indir}^\mathrm{2PN\,\ton{I}} = \rp{\nk\,\mu^2\,\ton{9 + 37\,e^2 + e^4}}{2\,c^4\,e^2\,a^2\,\ton{1-e^2}^2}.\lb{indir1}
\eqf
Note that \rfr{indir1} is formally singular in the limit $e\rightarrow 0$.

The second indirect contribution $\dot\omega_\mathrm{indir}^{\mathrm{2PN\,\ton{II}}}$ comes from the fact that, in general, when an extra-acceleration $\bds A$ like, e.g., \rfr{A1PN} enters the equations of motion, all its orbital parameters undergo instantaneous changes during an orbital period. Usually, in standard first order calculations in $A$, such generally slow variations are neglected by assuming the Keplerian elements as fixed to some fiducial values at a reference epoch $t_0$. Instead, accounting also for such changes yield further, indirect effects of the second order in $A$.
The resulting indirect integrated shift over one orbit of any of the orbital elements $\phi_i,\,i=1,\ldots 5$, where $\phi_1\doteq a,\,\phi_2\doteq e,\,\phi_3\doteq I,\,\phi_4\doteq\Omega,\phi_5\doteq\omega$,
can be calculated as
\eqi
\Delta\phi_i^{\ton{2}} = \sum_{j=1}^{5}\int_{f_0}^{f_0+2\uppi}\grf{\derp{\ton{\mathrm{d}\phi_i/\mathrm{d}f}}{\phi_j}}_\mathrm{K}\,
\Delta\phi_j\ton{f_0,\,f}^{\ton{1}}\,\mathrm{d}f,\,i=1,\ldots 5,\lb{grossa}
\eqf
where the superscript $\ton{2}$ indicates that the calculation is to the second order in $A$, $\grf{\ldots}_\mathrm{K}$ denotes that the content of the curly brackets has to be evaluated onto the unperturbed Keplerian ellipse, and $\Delta\phi_j\ton{f_0,\,f}^{\ton{1}},\, j=1,\ldots 5$ are the instantaneous shifts experienced by the orbital elements during the orbital revolution.
The latter ones are calculated as
\eqi
\Delta\phi_j\ton{f_0,\,f}^{\ton{1}} = \int_{f_0}^{f}\grf{\dert{\phi_j}{f^{'}}}_\mathrm{K}\,df^{'},\,j=1,\ldots 5,\lb{shi}
\eqf
where the superscript $\ton{1}$ indicates that the shifts of \rfr{shi} are to the first order in $A$.
From \citep{1959ForPh...7S..55T,1961mcm..book.....B,1970CeMec...2..369R,1979AN....300..313M}
\eqi
\dert\omega f = \rp{r^2}{\mu\,e}\grf{-\cos f\,A_r +\qua{1+\rp{r}{a\,\ton{1-e^2}}}\,\sin f\, A_\tau }-\cos I\,\dert\Omega f + \mathcal{O}\ton{A^2},\lb{dodf}
\eqf
valid to the first order in $A$ given, in the present case, by \rfr{A1PN}, and \rfrs{A1PNr}{A1PNn}, it turns out that, in the case of pericentre,  only the 1PN instantaneous shifts of $a$ and $e$ induced by \rfr{A1PN} are required. By recalling that
the Gauss equations for such orbital elements, to the first order in $A$, can be written as \citep{1959ForPh...7S..55T,1961mcm..book.....B,1970CeMec...2..369R,1979AN....300..313M}
\begin{align}
\dert a f & = \rp{2\,a\,r^2}{\mu\,\ton{1-e^2}}\qua{e\,A_r\,\sin f + \ton{\rp{p}{r}}\,A_\tau}+ \mathcal{O}\ton{A^2}, \\ \nonumber \\
\dert e f & = \rp{r^2}{\mu}\grf{A_r\,\sin f + \qua{\cos f + \rp{1}{e}\ton{1 - \rp{r}{a}}}\,A_\tau}+ \mathcal{O}\ton{A^2},
\end{align}
and that the radial, transverse, and out-of-plane components of \rfr{A1PN} are given by \rfrs{A1PNr}{A1PNn},
it is straightforward to obtain
\begin{align}
\Delta a\ton{f_0,\,f}^\mathrm{1PN} \lb{Da1PN} &= -\rp{2\,e\,\mu\,\ton{\cos f-\cos f_0}\,\qua{7 + 3\,e^2 + 5\,e\,\ton{\cos f+\cos f_0}}}{c^2\,\ton{1-e^2}^2}, \\ \nonumber \\
\Delta e\ton{f_0,\,f}^\mathrm{1PN} \lb{De1PN} &= \rp{\mu\,\ton{\cos f_0 - \cos f}\,\qua{3 + 7\,e^2 + 5\,e\,\ton{\cos f+\cos f_0}}}{c^2\,a\,\ton{1-e^2}}.
\end{align}
They agree with, e.g., Eq.\,(8.8.8) of \citet[p.\,331]{SoffelHan19}.
Their insertion in \rfr{grossa}, calculated for $i=5$ by means of \rfr{dodf}, yields
\eqi
\dot\omega_\mathrm{indir}^{\mathrm{2PN\,\ton{II}}} = -\rp{\nk\,\mu^2\,\grf{
9 - 87\,e^2 - 136\,e^4 + 19\,e^6 - 6\,e^3\,\qua{\ton{34 + 26\,e^2}\,\cos f_0 + 15\,e\,\cos 2 f_0}}}{2\,c^4\,e^2\,a^2\,\ton{1-e^2}^3}\lb{indir2}.
\eqf
Note that also \rfr{indir2} is formally singular in $e$; moreover, it depends on the initial value of the true anomaly $f_0$.

The indirect total 2PN precession $\dot\omega_\mathrm{indir}^\mathrm{2PN}$ of order $\mathcal{O}\ton{c^{-4}}$ is the sum of \rfr{indir1} and \rfr{indir2}; it reads
\eqi
\dot\omega^\mathrm{2PN}_\mathrm{indir} = \rp{\nk\,\mu^2\,\grf{5\,\ton{23 + 20\,e^2 - 4\,e^4} + 6\,e\,\qua{\ton{34+26\,e^2}\,\cos f_0 + 15\,e\,\cos 2f_0  } }}{2\,c^4\,a^2\,\ton{1-e^2}^3}.\lb{indir}
\eqf
It should be noticed that \rfr{indir} is not singular for $e\rightarrow 0$. On the other hand, \rfr{indir} is not univocally determined because of the presence of $f_0$. In Section\,\ref{numint}, we will confirm \rfr{indir} by numerically integrating the equations of motion for an arbitrary fictitious system.
%
%
\subsection{A numerical confirmation of the direct and indirect 2PN pericentre precessions}\lb{numint}
The direct 2PN precession of \rfr{dir2PN} was successfully confirmed by two numerical integrations of the equations of motion of, say, Mercury in the field of the Sun over $1\,\mathrm{century\,(cty)}$.

It is worthwhile recalling that the present-day level of accuracy in constraining any anomalous perihelion precession of such a planet with the most recent ephemerides, which all model the Solar system dynamics only up to the 1PN level in harmonic coordinates, may be at the level of $\upsigma_{\dot\omega}\simeq 8\,\mathrm{microarcseconds\,per\,century}\,\ton{\mu\mathrm{as\, cty}^{-1}}$, or, perhaps, $\simeq 10-50$ times worse; see the discussion in \citet{2019AJ....157..220I}, and references therein.

\begin{figure}[htb]
\begin{center}
\centerline{
\vbox{
\begin{tabular}{c}
\epsfysize= 8.0 cm\epsfbox{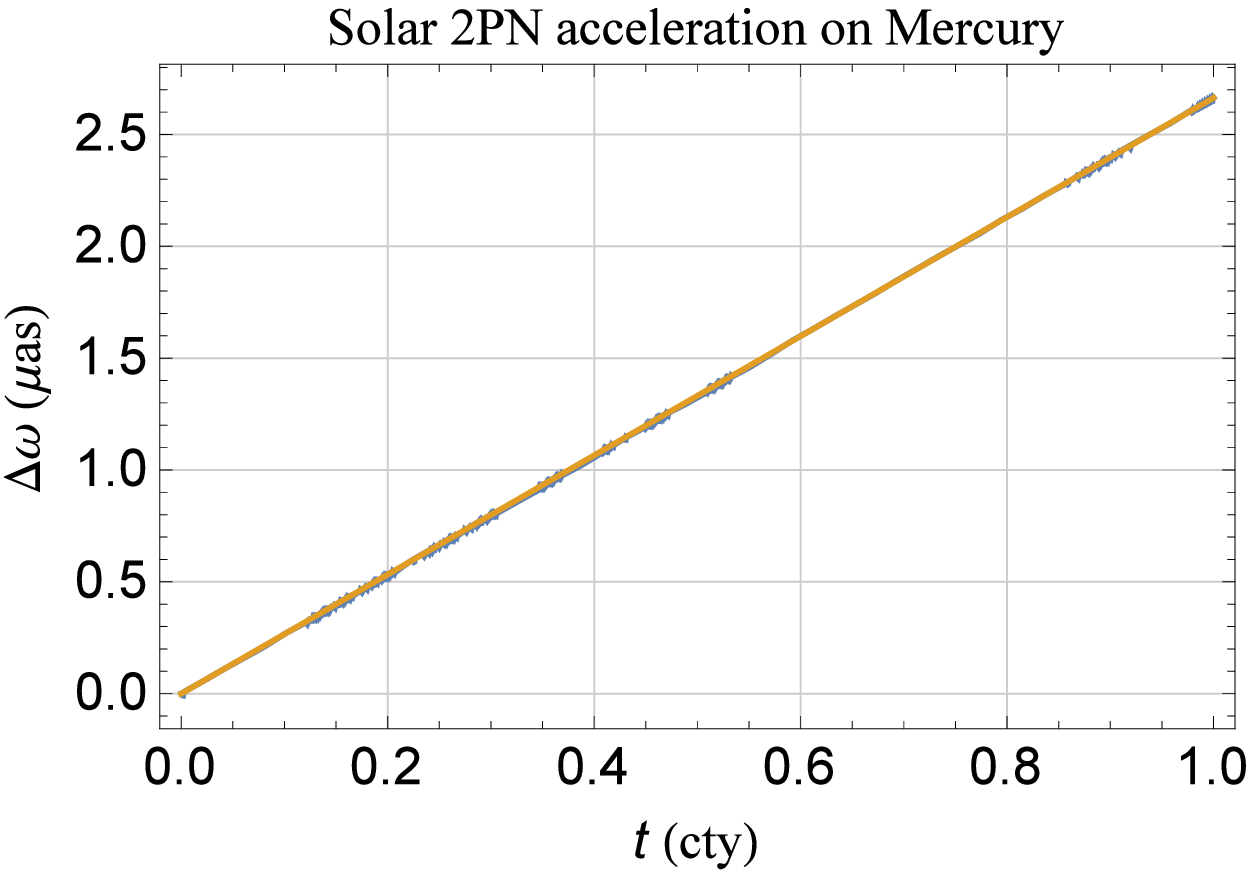}\\
\end{tabular}
}
}
\caption{
Numerically produced time series, in blue, of the 2PN evolution of the perihelion $\omega$ of Mercury over 1 cty calculated by numerically integrating the Hermean equations of motion, including the 2PN acceleration of \rfr{A2PN} in addition to the Newtonian monopole, in Cartesian rectangular coordinates along with the Gauss equations for all its Keplerian orbital elements. A superimposed linear fit, in yellow, to the numerically integrated time series of $\omega$ is displayed as well. Its slope of $2.6\,\mu\mathrm{as\,cty}^{-1}$ agrees with the value obtainable analytically by calculating \rfr{dir2PN} with the orbital parameters of Mercury.  The initial conditions were retrieved from the WEB interface HORIZONS by the NASA Jet Propulsion Labratory (JPL) which employs the same harmonic coordinates used in obtaining \rfr{A2PN} and \rfr{A1PN} to model the dynamics of the Solar system up to the 1PN level. The same plot, not displayed here, was obtained in a second numerical integration in which the Gauss equations were not included among the differential equations to be simultaneously solved.}\label{figura1}
\end{center}
\end{figure}

In the first run, we simultaneously integrated the Hermean equations of motion, including the Newtonian monopole and the 2PN acceleration of \rfr{A2PN}, in rectangular Cartesian coordinates along with the Gauss equations for all the Keplerian orbital elements over a time span 1 cty long starting from a set of initial conditions for the state vector of Mercury retrieved from the WEB interface HORIZONS, maintained by the NASA Jet Propulsion Laboratory (JPL). The resulting time series of the solution for $\omega\ton{t}$, in blue, is displayed in Figure\,\ref{figura1} along with a linear fit to it, in yellow.
The same plot was obtained in a second run in which the Gauss equations were not included in the numerical integration which was limited just to the equations of motion of Mercury in rectangular Cartesian coordinates, all the rest being the same as in the first run. Then, a time series for $\omega\ton{t}$ was straightforwardly computed from the solutions obtained for the Cartesian coordinates $x\ton{t},\,y\ton{t},\,z\ton{t}$ of the planet by means of the standard conversion formulas for the Keplerian orbital elements.
The resulting slope of the fitted linear trend amounts to $2.6\,\mu\mathrm{as\,cty}^{-1}$, in agreement with the first run and \rfr{dir2PN} calculated with the orbital parameters of Mercury. Interestingly, such a figure is only 3 times smaller than the previously quoted value of $\upsigma_{\dot\omega}$ which, however, as already remarked, may be optimistic by a factor of $\simeq 10-50$.

It should be noted that, at least in principle, the direct 2PN precession of \rfr{dir2PN} should be measurable since it is due to a distinct acceleration, i.e. \rfr{A2PN}, which may be suitably expressed in terms of a dedicated solve-for parameter to be estimated in a least-square sense in some covariance analyses.
Instead, the indirect precession of \rfr{indir}, since it comes from the 1PN acceleration of \rfr{A1PN} which is routinely modeled in the softwares of all the teams  currently producing the planetary ephemerides, may not be detectable as a separate effect with respect to the other 1PN features of motion.
Be that as it may, \rfr{indir} yields
\eqi
16\,\mu\mathrm{as\,cty}^{-1}\leq\dot\omega_\mathrm{indir}^\mathrm{2PN}\leq 33\,\mu\mathrm{as\,cty}^{-1}
\eqf
for $0\leq f_0< 360\,\mathrm{deg}$.

\begin{figure}
\begin{center}
\centerline{
\vbox{
\begin{tabular}{c}
\epsfysize= 7.0 cm\epsfbox{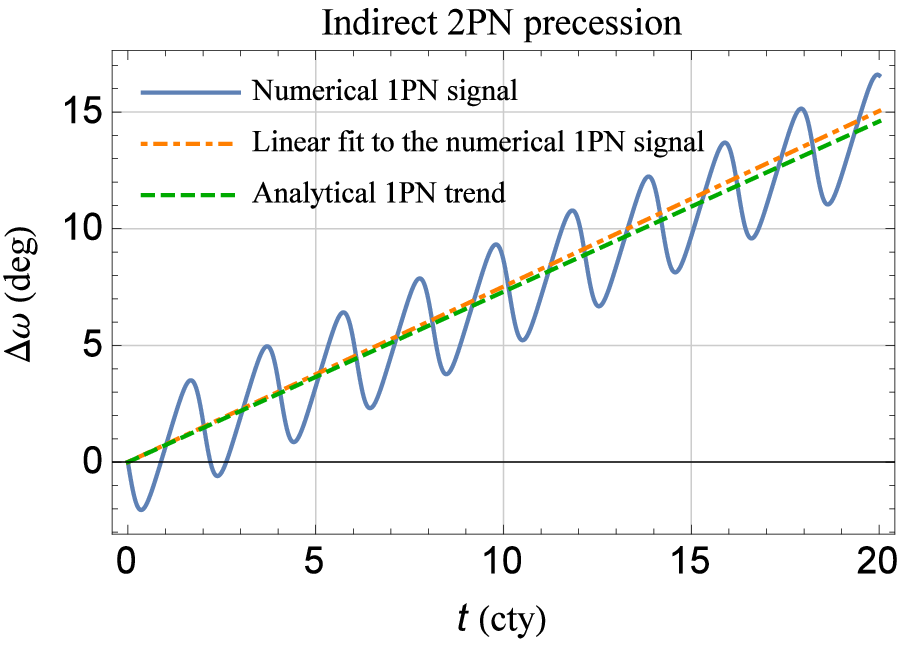}\\
\epsfysize= 7.0 cm\epsfbox{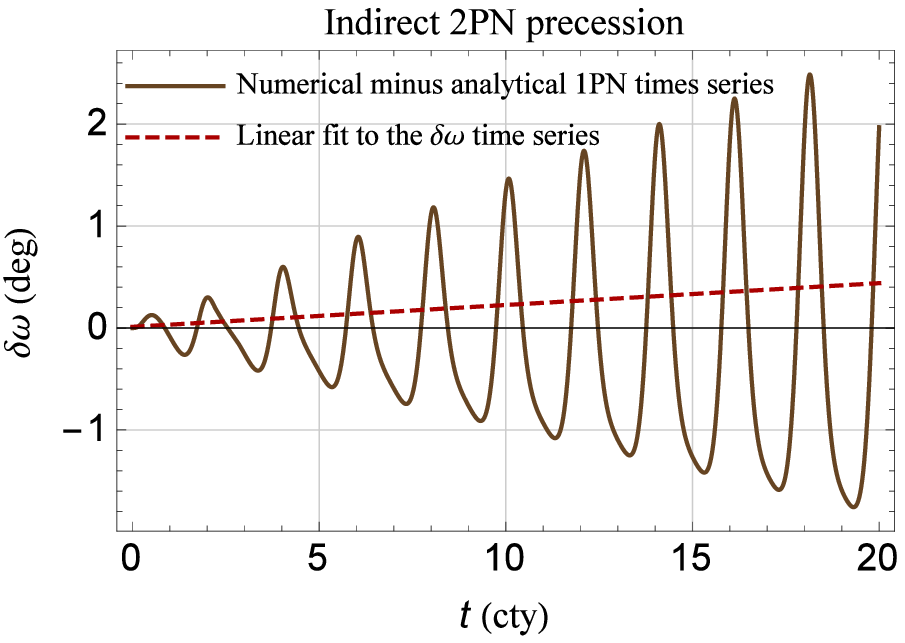}\\
\end{tabular}
}
}
\caption{Indirect 2PN pericenter precession in a fictitious scenario in which a test particle revolves around a primary with $M=10^{10}\,\mathrm{M}_\odot$ in $\Pb=2\,\mathrm{cty}$ around an elliptic orbit characterized by $e=0.095$. Upper panel: the continuous blue curve is the numerically produced time series of the overall (direct 1PN and indirect 2PN) pericenter shift $\Delta\omega(t)$ obtained by numerically integrating the equations of motion of the test particle over $10\,\Pb$ by including \textit{only} the 1PN acceleration of \rfr{A1PN}. It was obtained by taking the difference of the time series for $\omega$ computed from two integrations, with and without \rfr{A1PN}, sharing the same arbitrary initial conditions with, say, $f_0=0$. The dot-dashed orange straight line is a linear fit to $\Delta\omega(t)$, whose slope is $0.752\,\mathrm{deg\,cty}^{-1}$. The dashed green straight line is the analytical secular trend of the 1PN pericenter precession $\dot\omega^\mathrm{1PN}=0.730\,\mathrm{deg\,cty}^{-1}$. The difference between both the slopes of $0.022\,\mathrm{deg\,cty}^{-1}$ agrees just with the analytical prediction for the indirect 2PN precession of \rfr{indir} calculated with $f_0=0$. Lower panel: the continuous brown curve is the difference $\delta\omega$ between the continuous blue curve of the upper panel and the analytical 1PN time series for the pericenter of \rfr{peri1pn}, while the dashed red straight line is a linear fit to $\delta\omega$ with a slope of just $0.022\,\mathrm{deg\,cty}^{-1}$.
}\label{figura2}
\end{center}
\end{figure}

It is possible to numerically confirm our analytical findings also for the indirect 2PN precession in the following way. First of all, a straightforward numerical integration of the equations of motion of a fictitious restricted two-body system to the 1PN level, i.e. by accounting \textit{only} for the 1PN acceleration of \rfr{A1PN}, shows that the simple secular trend arising from the celebrated 1PN Einstein-like pericentre precession
\eqi
\dot\omega^\mathrm{1PN}=\rp{3\,\nk\,\mu}{c^2\,a\,\ton{1-e^2}}\lb{Eins}
\eqf
does \textit{not} match a linear fit to the time series obtained from the numerical integration. This is clearly shown in the upper panel of Figure\,\ref{figura2} obtained for, say, $f_0=0$. It turns out that such a feature lingers even by changing $f_0$ from a run to another. It is crucial to note that our analytical result for the indirect 2PN precession of \rfr{indir}, calculated with $f_0=0$, is able to fully explain the discrepancy between the slopes of the simple analytical 1PN trend due to \rfr{Eins} (dashed green line) and of the linear fit (dot-dashed orange line) to the numerically integrated overall signature (continuous blue curve) which, indeed, should include both the direct 1PN and the indirect 2PN effects altogether. It may be shown that it occurs for different values of $f_0$ as well. Such a feature is further confirmed by a more refined analysis, displayed in the lower panel of Figure\,\ref{figura2}, consisting of subtracting the well known analytical instantaneous time series of the 1PN change of $\omega$, given by \citep[Eq.\,(8.8.8),\,p.\,331]{SoffelHan19}
\begin{align}
\Delta\omega(f_0,\,f)^\mathrm{1PN} \nonumber & = \rp{\mu}{2\,c^2\,e\,a\,\ton{1-e^2}}\,\qua{6\,e\,\ton{f - f_0} + 2\,\ton{-3 + e^2}\,\sin f  - 5\,e\,\sin 2 f - \right.\\ \nonumber \\
&\left. - 2\,\ton{-3 + e^2}\,\sin f_0  + 5\,e\,\sin 2 f_0},\lb{peri1pn}
\end{align}
from the previously obtained numerical time series for the total (direct 1PN and indirect 2PN) time shift of the pericenter induced by the 1PN acceleration of \rfr{A1PN}. The resulting time series, obtained by expressing the true anomaly $f$ entering \rfr{peri1pn} as a function of time $t$ by means of \citep[p.~77]{1961mcm..book.....B}
\eqi
f\ton{t} = \mathcal{M}\ton{t} + 2\sum_{s = 1}^{s_\textrm{max}}\rp{1}{s}\grf{ J_s\ton{se} + \sum_{j = 1}^{j_\textrm{max}}\rp{\ton{1-\sqrt{1-e^2}}^j}{e^j}\qua{ J_{s-j}\ton{se} + J_{s+j}\ton{se}  }  }\sin s\mathcal{M}\ton{t}, \lb{fMt}
\eqf
where $\mathcal{M}=\nk\ton{t-t_0}+\mathcal{M}_0$ is the mean anomaly, $\mathcal{M}_0$ is the mean anomaly at epoch, $J_k\ton{se}$ is the Bessel function of the first kind of order $k$, and $s_\textrm{max},~j_\textrm{max}$ are some values of the summation indexes $s,~j$ adequate for the desired accuracy level,
is the continuous brown curve for $\delta\omega$ depicted in the lower panel of Figure\,\ref{figura2}. It can be noticed that it does not vanish, and a linear fit to it, represented by the dashed red line in the lower panel of Figure\,\ref{figura2}, returns just the same value as \rfr{indir}. Also in this case, it occurs by varying $f_0$.

\section{The case of a two-body system}\lb{mAmB}
\subsection{The direct pericentre precession due to the 2PN acceleration}\lb{dir2PNmAmB}
In the case of a two-body system made of two bodies $\mathrm{A},\,\mathrm{B}$ with masses $M_\mathrm{A},\,M_\mathrm{B}$, the 2PN acceleration of their relative motion is  \citetext{\citealp[see, e.g.,][Eq.\,(4.4.29),\,p.\,154]{1991ercm.book.....B}; \citealp[Eq.\,(2.2d),\,p.\,825]{1995PhRvD..52..821K};  \citealp[Eq.\,(B11),\,p.\,10]{PhysRevD.82.104031}}
\begin{align}
{\bds A}^\mathrm{2PN} \nonumber & = \rp{\mu}{c^4\,r^2}\grf{
\qua{\eta\,\ton{-3 + 4\,\eta}\,\mathrm{v}^4 + \rp{15}{8}\,\eta\,\ton{-1 + 3\,\eta}\,{\mathrm{v}}_r^4 +\,\eta\,\ton{\rp{9}{2} - 6\,\eta}\,\mathrm{v}^2\,{\mathrm{v}}_r^2 +\,\eta\,\ton{\rp{13}{2} - 2\,\eta}\,\rp{\mu}{r}\,\mathrm{v}^2 + \right.\right.\\ \nonumber \\
\nonumber &\left.\left. + \ton{2 + 25\,\eta + 2\,\eta^2}\,\rp{\mu}{r}\,{\mathrm{v}}_r^2 -\ton{9 + \rp{87}{4}\,\eta}\,\rp{\mu^2}{r^2}
}\,\bds{\hat{r}}
+\qua{
\eta\,\ton{\rp{15}{2} + 2\,\eta}\,\mathrm{v}^2 -\eta\,\ton{\rp{9}{2} + 3\,\eta}\,{\mathrm{v}}_r^2 - \right.\right.\\ \nonumber \\
&\left.\left. - \ton{2 + \rp{41}{2}\,\eta + 4\,\eta^2}\,\rp{\mu}{r}
}\,{\mathrm{v}}_r\,{\mathbf{v}}}.\lb{Mega2PN}
\end{align}
where $\mu\doteq GM,\,M\doteq M_\mathrm{A}+M_\mathrm{B}$, and $\eta\doteq M_\mathrm{A}\,M_\mathrm{B}/M^2$.

The direct 2PN precession $\dot\omega_\mathrm{dir}^\mathrm{2PN}$ of the pericentre of the relative motion of a two-body system can be straightforwardly computed from \rfr{Mega2PN} in the same fashion as for the point particle treated in Section\,\ref{dir2PNpp}.
The radial, transverse, and out-of-plane components of \rfr{Mega2PN} are
\begin{align}
\rp{64\,c^4\,\ton{1-e^2}^4}{a^5\,\nk^6\,\ton{1+e\,\cos f}^2}\,A_r^\mathrm{2PN} \nonumber & =
e^4\,\eta\,\ton{39 + 191\,\eta} + 16\,\qua{-36 +\,\eta\,\ton{-73 + 8\,\eta}} + \\ \nonumber \\
\nonumber & + 8\,e^2\,\qua{-36 +\,\eta\,\ton{-13 + 72\,\eta}} + \\ \nonumber \\
\nonumber & + 8\,e\,\grf{-144 +\,\eta\,\qua{-288 + 80\,\eta\,+ e^2\,\ton{13 + 92\,\eta}}}\,\cos f + \\ \nonumber \\
\nonumber & + e^2\,\grf{4\,\qua{-72 +\,\eta\,\ton{-298 + 144\,\eta\,+
e^2\,\ton{-45 + 11\,\eta}}}\,\cos 2 f + \right. \\ \nonumber \\
\nonumber &\left. + e\,\eta\,\qua{8\,\ton{-57 + 20\,\eta}\,\cos 3 f + 3\,e\,\ton{-17 + 7\,\eta}\,\cos 4 f}}\lb{arr}
, \\ \nonumber \\
A_\tau^\mathrm{2PN} & = -\rp{a^5\,e\,\nk^6\,\ton{1+e\,\cos f}^3\,\sin f}{2\,c^4\,\ton{1-e^2}^4}\grf{
4 + \eta\,\qua{26 + 4\,\eta - e^2\,\ton{15 + 4 \eta}} + \right.\\ \nonumber \\
&\left. + e\,\ton{4 + 11\,\eta}\,\cos f + 3\,e^2\,\eta\,\ton{3 + 2 \eta}\,\sin^2 f
}, \\ \nonumber \\
A_\nu^\mathrm{2PN} & = 0\lb{ann};
\end{align}
they reduce to \rfrs{Ar2PN}{An2PN} in the point particle limit, i.e. for $\eta\rightarrow 0$.
By averaging the right-hand sides of \rfrs{gaus1}{gaus2}, calculated with \rfrs{arr}{ann}, with the first term of \rfr{dtdf} one finally obtains
\eqi
\dot\omega_\mathrm{dir}^\mathrm{2PN} = \rp{\nk\,\mu^2\grf{
e^2\,\qua{-2 + 3\,\ton{7 - 16\,\eta}\,\eta} + 8\,\qua{7 + \ton{5 - 7\,\eta}\,\eta} }}{8\,c^4\,a^2\,\ton{1-e^2}^2}\lb{Dir2PN}.
\eqf
\rfr{Dir2PN} reduces to \rfr{dir2PN} for $\eta\rightarrow 0$.

For the double pulsar PSR J0737−3039A/B, characterized by \citep{2006Sci...314...97K} $M_\mathrm{A}=1.3381\,\mathrm{M}_\odot,\,M_\mathrm{B}=1.2489\,\mathrm{M}_\odot,\,\eta=0.249,\,M=2.58708\,\mathrm{M}_\odot,\,a=878960\,\mathrm{km},\,e=0.0877,\,\Pb=0.10\,\mathrm{d}$, \rfr{Dir2PN} yields
\eqi
\dot\omega_\mathrm{dir}^\mathrm{2PN} = 0.00019\,\mathrm{deg\,yr}^{-1}.\lb{doppia}
\eqf
The current accuracy in measuring the periastron precession of the double pulsar is \citep{2006Sci...314...97K}
\eqi
\upsigma_{\dot\omega}=0.00068\,\mathrm{deg\,yr}^{-1}.
\eqf
An accuracy level of the order of \rfr{doppia} should be reached in the forthcoming years thanks to new telescopes \citep{2018mgm..conf.1860K}.
For the historical binary pulsar PSR B1913+16,  whose relevant physical and orbital parameters are \citep{2010ApJ...722.1030W} $M_\mathrm{A}=1.4398\,\mathrm{M}_\odot,\,M_\mathrm{B}=1.3886\,\mathrm{M}_\odot,\,\eta=0.249,\,M=2.8284\,\mathrm{M}_\odot,\,a=1.949\times 10^6\,\mathrm{km},\,e=0.6171334,\,\Pb=0.32\,\mathrm{d}$ , \rfr{Dir2PN} returns
\eqi
\dot\omega_\mathrm{dir}^\mathrm{2PN} = 0.000038\,\mathrm{deg\,yr}^{-1},\lb{binaria}
\eqf
while the most recent determination of its periastron rate is accurate to \citep{2010ApJ...722.1030W}
\eqi
\upsigma_{\dot\omega}=0.000005\,\mathrm{deg\,yr}^{-1}.
\eqf
For the supermassive binary black hole in OJ 287, whose relevant orbital parameters are \citep{2018ApJ...866...11D} $M_\mathrm{A}=18438\times 10^6\,\mathrm{M}_\odot,\,M_\mathrm{B}=150.13\times 10^6\,\mathrm{M}_\odot,\,\Pb=12.06\,\mathrm{yr},\,e=0.657$, \rfr{Dir2PN} predicts a direct 2PN perinigricon\footnote{It is one of the possible names which can be attributed  to the pericentre when black holes are involved \citep{2002Natur.419..694S}. It comes from the Latin word \virg{\textit{niger}}, meaning \virg{black}.} precession as large as $\dot\omega_\mathrm{dir}^\mathrm{2PN}=11.0\,\mathrm{deg\,cty}^{-1}$, a remarkable fraction of the 1PN rate of change
\eqi
\dot\omega^\mathrm{1PN}=\rp{3\,\nk\,\mu}{c^2\,a\,\ton{1-e^2}}=206.8\,\mathrm{deg\,cty}^{-1}\lb{dodt1PN}
\eqf
corresponding to a shift per orbit
\eqi
\Delta\omega^\mathrm{1PN}=24.9\,\mathrm{deg}.
\eqf
\subsection{The indirect pericentre precession due to the 1PN acceleration}\lb{indir2PNmAmB}
The indirect precession due to the 1PN acceleration \citetext{\citealp[see, e.g.,][Eq.\,(4.4.28),\,p.\,154]{1991ercm.book.....B}; \citealp[Eq.\,(A2.6),\,p.\,166]{Sof89}; \citealp[Eq.\,(10.3.7),\,p.\,381]{SoffelHan19}}
\eqi
{\bds A}^\mathrm{1PN} = \rp{\mu}{c^2\,r^2}\grf{\qua{\ton{4+2\,\eta}\,\rp{\mu}{r} + \rp{3}{2}\,\eta\,{\mathrm{v}}_r^2-\ton{1+3\,\eta}\,\mathrm{v}^2}\,\bds{\hat{r}}+ \ton{4-2\,\eta}\,{\mathrm{v}}_r\,{\mathbf{v}}}\lb{a1pn}
\eqf
can be calculated as in the point particle case treated in Section\,\ref{indir2PNpp}.

The radial, transverse, and out-of-plane components of \rfr{a1pn} are
\begin{align}
A_r^\mathrm{1PN} \lb{R1PN}& = \rp{\mu^2\,\ton{1+e\,\cos f}^2\,\qua{e^2\,\ton{4 - 13\,\eta} - 4\,\ton{-3 +\,\eta} + 8\,e\,\ton{1 - 2\,\eta}\,\cos f + e^2\,\ton{-8 +\,\eta}\,\cos 2f}}{4\,c^2\,a^3\,\ton{1-e^2}^3}, \\ \nonumber \\
A_\tau^\mathrm{1PN} \lb{T1PN}& = \rp{2\,e\,\mu^2\,\ton{1+e\,\cos f}^3\,\ton{2-\eta}\,\sin f}{c^2\,a^3\,\ton{1-e^2}^3}, \\ \nonumber \\
A_\nu^\mathrm{1PN} \lb{N1PN}& = 0.
\end{align}
\rfrs{R1PN}{N1PN}, which agree with Equations\,(A2.77a)-(A2.77c) of \citet[p.\,178]{Sof89}, reduce to \rfrs{A1PNr}{A1PNn} for $\eta\rightarrow 0$.

The indirect precession $\dot\omega_\mathrm{indir}^\mathrm{2PN\,\ton{I}}$ due to the second and third terms of \rfr{dtdf} turns out to be
\eqi
\dot\omega_\mathrm{indir}^\mathrm{2PN\,\ton{I}} = \rp{\nk\,\mu^2\grf{32\,\ton{-3 +\,\eta}^2 + 8\,e^2\,\qua{148 + 5\,\eta\,\ton{-43 + 17\,\eta}} + e^4\,\qua{32 + 3\,\eta\,\ton{56 + 75\,\eta}}}}{64\,c^4\,e^2\,a^2\,\ton{1-e^2}^2}.\lb{Indir1}
\eqf
\Rfr{Indir1} reduces to \rfr{indir1} in the point particle limit.

The 1PN instantaneous shifts of $a$ and $e$ induced by \rfr{a1pn} are
\begin{align}
\Delta a\ton{f_0,\,f}^\mathrm{1PN} \nonumber \lb{da1pn}& = \rp{e\,\mu\ton{\cos f-\cos f_0}}{2\,c^2\,\ton{1-e^2}^2}\grf{
4\,\qua{-7 + 3\,\eta + e^2\,\ton{-3 + 4\,\eta}} + \right.\\ \nonumber \\
\nonumber &\left. + e\,\qua{e\,\eta\,\cos 2f + 4\,\ton{-5 + 4\,\eta}\,\cos f_0 + 2\,\cos f \ton{-10 + 8\,\eta\,+ e\,\eta\,\cos f_0} + \right.\right.\\ \nonumber \\
&\left.\left. + e\,\eta\,\cos 2f_0}
}, \\ \nonumber \\
\Delta e\ton{f_0,\,f}^\mathrm{1PN} \nonumber \lb{de1pn}& = \rp{\mu\ton{\cos f-\cos f_0}}{4\,c^2\,a\,\ton{1-e^2}}\grf{
4\,\qua{-3 + \eta + e^2\,\ton{-7 + 6\,\eta}} + \right.\\ \nonumber \\
\nonumber &\left. + e\,\qua{e\,\eta\,\cos 2f + 4\,\ton{-5 + 4\,\eta}\,\cos f_0 + 2\,\cos f\,\ton{-10 + 8\,\eta + e\,\eta\,\cos f_0} + \right.\right.\\ \nonumber \\
&\left.\left. + e\,\eta\,\cos 2f_0}
}.
\end{align}
They agree with Equations\,(A2.78b)-(A2.78c) of \citet[p.\,178]{Sof89}, and reduce to \rfrs{Da1PN}{De1PN} in the limit $\eta\rightarrow 0$. \Rfrs{da1pn}{de1pn} allow to compute the other indirect contribution $\dot\omega^\mathrm{2PN\,\ton{II}}_\mathrm{indir}$ to the 2PN precession, which reads
\begin{align}
-\rp{64\,c^4\,e^2\,a^2\,\ton{1-e^2}^3}{\nk\,\mu^2}\,\dot\omega^\mathrm{2PN\,\ton{II}}_\mathrm{indir} \nonumber \lb{Indir2} &=
32\,\ton{-3 +\,\eta}^2 - 8\,e^2\,\ton{-3 +\,\eta}\,\ton{-116 + 47\,\eta} + \\ \nonumber \\
\nonumber &+ e^4\,\qua{-4352 + \ton{10664 - 4183\,\eta}\,\eta} +\\ \nonumber \\
\nonumber & + e^6\,\qua{608 + 3\,\ton{304 - 601\,\eta}\,\eta} + \\ \nonumber \\
\nonumber & + 48\,e^3\,\grf{\qua{8\,\ton{-17 + 7\,\eta} + e^2\,\ton{-104 + 109\,\eta}}\,\cos f_0  + \right. \\ \nonumber \\
&\left. + 3\,e\,\qua{4\,\ton{-5 + 4\,\eta}\,\cos 2 f_0  +  e\,\eta\,\cos 3 f_0}}.
\end{align}

The sum of \rfr{Indir1} and \rfr{Indir2}, which reduces to \rfr{indir2} for $\eta\rightarrow 0$, yields the total indirect 2PN precession, which is
\begin{align}
-\rp{32\,c^4\,a^2\,\ton{1-e^2}^3}{\nk\,\mu^2}\,\dot\omega_\mathrm{indir}^\mathrm{2PN} \nonumber & = e^4\,\ton{320 + 540\,\eta - 789\,\eta^2} - 16\,\qua{115 + 16\,\eta\,\ton{-7 + 2\,\eta}} - \\ \nonumber \\
\nonumber & - 4\,e^2\,\qua{400 + \eta\,\ton{-1097 + 466\,\eta}} + \\ \nonumber \\
\nonumber & + 24\,e\,\grf{\qua{8\,\ton{-17 + 7\,\eta} + e^2\,\ton{-104 + 109\,\eta}}\,\cos f_0 + \right.\\ \nonumber \\
&\left. + 3\,e\,\qua{4\,\ton{-5 + 4\,\eta}\,\cos 2 f_0 + e\,\eta\,\cos 3 f_0}}.\lb{Indir}
\end{align}
\Rfr{Indir} agrees with \rfr{indir} in the point particle limit.

According to \rfr{Indir}, the indirect periastron precession of PSR J07373039A/B lies in the range
\eqi
0.00092\,\mathrm{deg\,yr}^{-1}\leq\dot\omega^\mathrm{2PN}_\mathrm{indir}\leq 0.00132\,\mathrm{deg\,yr}^{-1}
\eqf
for $0\leq f_0<360\,\mathrm{deg}$. If summed to the direct precession of \rfr{doppia}, such a result would bring the total 2PN periastron precession of the double pulsar in the realm of measurability independently of $f_0$.
For the binary pulsar PSR B1913+16, the indirect 2PN precession of \rfr{Indir} is
\eqi
-0.000048\,\mathrm{deg\,yr}^{-1}\leq\dot\omega_\mathrm{indir}^\mathrm{2PN}\leq 0.001052\,\mathrm{deg\,yr}^{-1}\lb{nihil}
\eqf
for $0\leq f_0<360\,\mathrm{deg}$.
This implies that, for certain values of $f_0$, \rfr{nihil} may cancel the direct precession of \rfr{binaria}, thus making a potential measurement of the 2PN orbital effect unmeasurable.
For OJ 287, \rfr{Indir} yields an indirect 2PN perinigricon precession ranging from a maximum of $516\,\mathrm{deg\,cty}^{-1}$ to a minimum of $20\,\mathrm{deg\,cty}^{-1}$. It is a remarkable result in view of \rfr{dodt1PN}.
\section{A comparison with other works}\lb{parago}
To the knowledge of the present author, the only other work in the literature making use of the method of the variation of constants and the Gauss equations is \citet{1994ARep...38..104K}.  As we will show, their result is incorrect because of the treatment of what are dubbed here as indirect effects.

Equation\,(5.2) of \citet{1994ARep...38..104K}, which we reproduce here to the benefit of the reader, is their main result. It is the \textit{total} 2PN pericenter shift per orbit, in units of $2\uppi$, written in terms of the constants of integration $k_1,\,k_2$ of the solutions of the Gauss equations for the semimajor axis and the eccentricity to the 1PN level. In our notation\footnote{In \citet{1994ARep...38..104K}, it is $k_1\rightarrow a_0,\,k_2\rightarrow e_0$.}, it is,  in the test particle case,
\eqi
\rp{\Delta\omega^\mathrm{2PN}_\mathrm{tot}}{2\uppi}=\rp{3\,\mu}{c^2\,k_1\,\ton{1-k_2^2}}\qua{1+\rp{3\,\mu}{4\,c^2\,k_1\,\ton{1-k_2^2}}-\rp{\mu}{4\,c^2\,k_1}}.\lb{kop1}
\eqf
Since the constants of integrations $k_1,\,k_2$ entering \rfr{kop1} are determined with the initial conditions at $t=t_0$, they contain explicitly $f_0$; thus, Equation\,(5.2) of \citet{1994ARep...38..104K} actually does depend on the latter one, contrary to what, at first glance, someone could argue, perhaps mislead by the notation used by \citet{1994ARep...38..104K} for $k_1,\,k_2$.
%
%
%
%
%
%
By retrieving the explicit expression of $k_1,\,k_2$ from \rfrs{Da1PN}{De1PN}
\begin{align}
k_1 \lb{k1} &= a + \rp{e\,\mu\,\qua{\ton{14+6\,e^2}\,\cos f_0 +e\,\ton{4+5\cos 2f_0}  }}{c^2\,\ton{1-e^2}^2}, \\ \nonumber \\
k_2 \lb{k2} &= e + \rp{\mu\,\qua{\ton{6+14\,e^2}\,\cos f_0 +e\,\ton{2+5\cos 2f_0} }}{2\,c^2\,a\,\ton{1-e^2}},
\end{align}
where $a$ and $e$ entering \rfrs{k1}{k2} are intended as the Keplerian values of the unperturbed case,
 Equation\,(5.2) of \citet{1994ARep...38..104K} can be finally cast into the form
\eqi
\rp{\Delta\omega^\mathrm{2PN}_\mathrm{tot}}{2\uppi}=\rp{3\,\mu^2\ton{2+e^2-32\,e^2\,\cos f_0}}{4\,c^4\,a^2\,\ton{1-e^2}^2},\lb{koppete}
\eqf
which does \textit{not} agree with the corresponding expression for $\Delta\omega_\mathrm{tot}^\mathrm{2PN}/2\uppi$ obtainable from the sum of our \rfr{dir2PN} and \rfr{indir} by taking its ratio to $\nk$.

From what can be deduced from the description of the method followed by \citet{1994ARep...38..104K}, the indirect effect corresponding to our $\dot\omega_\mathrm{indir}^\mathrm{2PN\,(II)}$ arises from the replacement $a\rightarrow a+\Delta a\ton{f_0,\,f}^\mathrm{1PN},\,e\rightarrow e+\Delta e\ton{f_0,\,f}^\mathrm{1PN}$ in\footnote{It is done in the first term of Equation\,(5.1) of \citet{1994ARep...38..104K} when  Equation\,(3.6) of \citet{1994ARep...38..104K} for $d\omega/dt$ is calculated to the 1PN level.} \rfr{dodf}, in a series expansion of it in powers of $c^{-1}$ to the order $c^{-4}$, and in an integration of the resulting expression from $f_0$ to $f_0+2\uppi$. The result, not explicitly shown by \citet{1994ARep...38..104K}, is
\eqi
\rp{\Delta\omega_\mathrm{indir}^\mathrm{2PN\,(II)}}{2\uppi} = \rp{\mu^2\,\ton{-9 - 48\,e^2 + e^4 - 48\,e^3\,\cos f_0}}{2\,c^4\,a\,e^2\,\ton{1-e^2}^2},\lb{kop2}
\eqf
which does not agree with the corresponding expression from our \rfr{indir2} for $\dot\omega_\mathrm{indir}^\mathrm{2PN\,(II)}$.
Instead, it seems that the other two contributions arising from Equation\,(5.1) of \citet{1994ARep...38..104K}, despite not explicitly displayed by \citet{1994ARep...38..104K}, agree with the corresponding shifts from our \rfr{dir2PN} and \rfr{indir1} because their sum with \rfr{kop2} yields just \rfr{koppete}. In particular, the fractional 2PN advance per orbit, which should come from the first term of Equation\,(5.1) of \citet{1994ARep...38..104K} calculated with ${\bds A}^\mathrm{2PN}$ onto a reference Keplerian ellipse, is not shown; nonetheless,  from the description of the calculational method by \citet{1994ARep...38..104K}, one may expect that it agrees with our \rfr{dir2PN}. Moreover, a direct calculation confirms that the second term of Equation\,(5.1) of \citet{1994ARep...38..104K} yields just the shift corresponding to our \rfr{indir1} for $\dot\omega_\mathrm{indir}^\mathrm{2PN\,(I)}$.
Thus, it can be inferred that the total indirect 2PN pericentre precession of \citet{1994ARep...38..104K} can be cast into the form
\eqi
\dot\omega_\mathrm{indir}^\mathrm{2PN} = \rp{\nk\,\mu^2\ton{-11 + 2\,e^2 - 48\,e\,\cos f_0}}{2\,c^4\,a^2\,\ton{1-e^2}^2}\lb{indirkop}.
\eqf
It neatly disagrees with our numerical results of Section\,\ref{numint} since, for the fictitious system treated in Figure\,\ref{figura2}, \rfr{indirkop} provides a slope as little as $-0.00255\,\mathrm{deg\,cty}^{-1}$.

It may be interesting to make a comparison of our results also with the seminal results by \citet{1988NCimB.101..127D}, despite they did not use the Gauss equations. \citet{1988NCimB.101..127D}, following the example by \citet{1971ctf..book.....L}, started from the Hamiltonian of the binary system in Arnowitt-Deser-Misner (ADM) coordinates  \citep{1960PhRv..120..313A} and adopted the Hamilton-Jacobi method.
As far as the 2PN pericentre precession of a system of two mass monopoles is concerned, their main result is Equation\,(3.12)
\eqi
\rp{\Delta\omega_\mathrm{tot}^\mathrm{2PN}}{2\uppi} = \rp{3}{c^2\,h^2}\,\qua{1+\ton{\rp{5}{2}-\eta}\rp{E}{c^2} +\ton{\rp{35}{4} -\rp{5}{2}\eta }\rp{1}{c^2\,h^2}, },\lb{DS1}
\eqf
where $h$ and $E$ are the coordinate-invariant, reduced orbital angular momentum and energy, respectively.
Its translation in terms of the parameters of the Damour-Deruelle (DD) parametrization \citep{1985AIHS...43..107D} is given by Equation\,(5.18) of \citet{1988NCimB.101..127D}
\begin{align}
\rp{\Delta\omega_\mathrm{tot}^\mathrm{2PN}}{2\uppi} \nonumber  &=\rp{3\ton{\mu\,n}^{2/3}}{c^2\,\ton{1-e^2_t}}\,
\qua{1 + \rp{\ton{\mu\,n}^{2/3}}{c^2\,\ton{1-e^2_t}}\,\ton{\rp{39}{4}\,x_\mathrm{A}^2 +\rp{27}{4}\,x_\mathrm{B}^2 + 15\,x_\mathrm{A}\,x_\mathrm{B}} -\right.\\ \nonumber \\
&-\left. \rp{\ton{\mu\,n}^{2/3}}{c^2}\,\ton{\rp{13}{4}\,x_\mathrm{A}^2 +\rp{1}{4}\,x_\mathrm{B}^2 + \rp{13}{3}\,x_\mathrm{A}\,x_\mathrm{B}} }\lb{DS2}
\end{align}
where
$n$ is the PN mean motion \citep[Equation\,(3.6d)]{1985AIHS...43..107D}
\eqi
n=\rp{\ton{-2\,E}^{3/2}}{\mu}\,\qua{1-\rp{E}{4\,c^2}\,\ton{\eta-15}},\lb{mmot}
\eqf
\eqi
x_\mathrm{A}\doteq \rp{M_\mathrm{A}}{M},\,x_\mathrm{B}\doteq \rp{M_\mathrm{B}}{M}=1-x_\mathrm{A},
\eqf
and $e_t$ is one of the DD parameters \citep{1985AIHS...43..107D}.
Expressing \rfr{DS1} in terms of the osculating Keplerian orbital elements can be made in the following two steps. First, $E,\,h$ are to be written in terms of the DD parameters $a_r,\,e_r$ by inverting Equations\,(3.6a) and Equation\,(3.6b) of \citet{1985AIHS...43..107D}.
\begin{align}
E &= -\rp{\mu}{2\,a_r}\rp{1}{\qua{1+\rp{\mu}{4\,c^2\,a_r}\,\ton{7-\eta} }}, \\ \nonumber \\
h^2 & = \rp{a_r\,\ton{1-e_r^2}+\rp{\mu}{2\,c^2}\qua{19 + e_r^2\ton{-7 + \eta} - 3\,\eta}-\rp{\mu^2}{16\,c^4\,a_r}\,\qua{-577 + e_r^2\,\ton{-7 + \eta}^2 + \ton{246 - 25\,\eta}\,\eta}    }{\mu\qua{1 + \rp{\mu}{2\,c^2\,a_r}\,\ton{11-3\,\eta} }}.
\end{align}

Then, Equations\,(28)\,to\,(29) of \citet{1994ApJ...427..951K},
which in general relativity are
\begin{align}
a_r & = \rp{a\,\ton{1 - e^2}^2 - da_0\,\ton{1 - e^2}^2 - \rp{\mu}{c^2}\,\qua{-3 + \eta + e^2\,\ton{-13 + e^2 + 7\,\eta + 2\,e^2\,\eta}}}{\ton{1 - e^2}^2},\\ \nonumber \\
e_r & =\rp{-2\,a\,\ton{de_0-e}\,\ton{-1+e^2}+\rp{e\,\mu}{c^2}\,\qua{-17 + 6\,\eta + e^2\,\ton{2 + 4\,\eta}}  }{2\,a\,\,\ton{-1+e^2}},
\end{align}
with the aid of Equation\,(14) and Equation\,(16) of \citet{1994ApJ...427..951K},
whose general relativistic expressions are
\begin{align}
da_0 \lb{da0} & = \rp{e\,\mu\,\grf{\qua{8\,\ton{-7 + 3\,\eta} + e^2\,\ton{-24 + 31\,\eta}}\,\cos f_0  + e\,\qua{4\,\ton{-5 + 4\,\eta}\,\cos 2 f_0  + e\,\eta\,\cos 3 f_0}}}{4\,c^2\,\ton{1-e^2}^2}, \\ \nonumber \\
de_0 \lb{de0} & = -\rp{\mu\,\grf{\qua{8\,\ton{-3 +\eta} + e^2\,\ton{-56 + 47\,\eta}}\,\cos f_0 + e\,\qua{4\,\ton{-5 + 4\,\eta}\,\cos 2 f_0 + e\,\eta\,\cos 3 f_0}}}{8\,c^2\,a\,\ton{-1+e^2}},
\end{align}
are used to express $a_r,\,e_r$ as functions of the osculating Keplerian elements $a,\,e$.
We obtain for $a_r\ton{a,\,e},\,e_r\ton{a,\,e}$
\begin{align}
4\,\ton{1-e^2}^2\,a_r \nonumber \lb{aerre}& = 4\,\grf{a\,\ton{1 - e^2}^2 - \rp{\mu}{c^2}\,\qua{-3 + \eta + e^4\,\ton{1 + 2\,\eta} + e^2\,\ton{-13 + 7\,\eta}}} + \\ \nonumber \\
\nonumber & + e\,\rp{\mu}{c^2}\,\grf{\qua{56 + e^2\,\ton{24 - 31\,\eta} - 24\,\eta}\,\cos f_0 + \right. \\ \nonumber \\
&+\left. e\,\qua{4\,\ton{5 - 4\,\eta}\,\cos 2 f_0 - e\,\eta\,\cos 3 f_0}}, \\ \nonumber \\
8\,a\,\ton{-1+e^2}\,e_r \nonumber \lb{eerre}& = 4\,e\,\grf{2\,a\,\ton{-1 + e^2} + \rp{\mu}{c^2}\,\qua{-17 + 6\,\eta + e^2\,\ton{2 + 4\,\eta}}} + \\ \nonumber \\
\nonumber &+ \rp{\mu}{c^2}\,\grf{\qua{8\,\ton{-3 + \eta} + e^2\,\ton{-56 + 47\,\eta}}\,\cos f_0  + \right. \\ \nonumber \\
&+\left. e\,\qua{4\,\ton{-5 + 4\,\eta}\,\cos 2 f_0  + e\,\eta\,\cos 3 f_0}}.
\end{align}
Finally, an expansion of the obtained expression in powers of $c^{-1}$ to the 2PN level yields, in the point particle limit, \rfr{koppete} which, as already noted, is incorrect.
On the other hand, \rfr{DS1} and \rfr{DS2} seem to be mutually inconsistent since their expressions in terms of $a,\,e$ do not even agree each other. Indeed,
by using \rfr{mmot} and \rfrs{aerre}{eerre}, Equation\,(30) of \citet{1994ApJ...427..951K}, which, in general relativity, reads
\eqi
e_t = \rp{-2\,a\,\ton{de_0-e}\,\ton{-1+e^2} +\rp{e\,\mu}{c^2}\,\qua{3\,\ton{-3 + \eta} + e^2\ton{-6 + 7\,\eta}} }{2\,a\,\ton{-1+e^2}},
\eqf
and \rfr{de0} to express $e_t$ in terms of $a,\,e$
\begin{align}
8\,a\,\ton{-1 + e^2}\,e_t \nonumber &= 4\,e\,\grf{2\,a\,\ton{-1 + e^2} + \rp{\mu}{c^2}\,\qua{3\,\ton{-3 + \eta} + e^2\,\ton{-6 + 7\,\eta}}} + \\ \nonumber \\
\nonumber &+ \rp{\mu}{c^2}\,\grf{\qua{8\,\ton{-3 + \eta} + e^2\,\ton{-56 + 47\,\eta}}\,\cos f_0 + \right.\\ \nonumber \\
&+\left. e\,\qua{4\,\ton{-5 + 4\,\eta}\,\cos 2 f_0 + e\,\eta \,\cos 3 f_0}},
\end{align}
one obtains, in the limit $\eta\rightarrow 0$,
\eqi
\rp{\Delta\omega_\mathrm{tot}^\mathrm{2PN}}{2\uppi} = \rp{3\,\mu^2\,\ton{2-3\,e^2 -32\,e\cos f_0 }}{4\,c^4\,a^2\,\ton{1-e^2}^2},\lb{koppete2}
\eqf
which disagrees even with \rfr{koppete} itself. By expanding \rfr{koppete} and \rfr{koppete2} in powers of $e$, it turns out that their disagreement is at the order $\mathcal{O}\ton{e^2}$.

\section{Summary and conclusions}\lb{fine}
We analytically worked out the 2PN secular pericentre precession $\dot\omega^\mathrm{2PN}$ of both a test particle orbiting a static central body and a full two-body system made of a pair of comparable non-rotating monopole masses with the method of variation of orbital elements.

We, first, calculated the direct precession $\dot\omega_\mathrm{dir}^\mathrm{2PN}$ induced by the 2PN acceleration entering the equations of motion written in harmonic coordinates. Two different numerical integrations of the equations of motion of a point particle confirmed our analytical results. For Mercury moving in the field of Sun, it is $\dot\omega_\mathrm{dir}^\mathrm{2PN}=2.6\,\mu\mathrm{as\,cty}^{-1}$. It is just 3 times smaller than the present-day \textit{formal} accuracy $\upsigma_{\dot\omega}=8\,\mu\mathrm{as\,cty}^{-1}$ in constraining any unmodelled effect in the Hermean perihelion rate with the latest planetary ephemerides, although $\upsigma_{\dot\omega}$ may be realistically up to $\simeq 10-50$ times worse. In the case of the binary pulsar PSR B1913+16, the direct 2PN periastron rate is
$\dot\omega_\mathrm{dir}^\mathrm{2PN} = 0.000038\,\mathrm{deg\,yr}^{-1}$, to be compared with the most recent determination of its periastron rate
$\upsigma_{\dot\omega}=0.000005\,\mathrm{deg\,yr}^{-1}$, while for the  double pulsar PSR J0737−3039A/B one has
$\dot\omega_\mathrm{dir}^\mathrm{2PN} = 0.00019\,\mathrm{deg\,yr}^{-1}$ and
$\upsigma_{\dot\omega}=0.00068\,\mathrm{deg\,yr}^{-1}$. The direct 2PN perinigricon precession of the supermassive binary black hole in OJ 287 amounts to $11\,\mathrm{deg\,cty}^{-1}$.

Then, we computed also the indirect 2PN pericentre precession $\dot\omega_\mathrm{indir}^\mathrm{2PN}$ arising from the fact that the 1PN acceleration actually changes instantaneously the semimajor axis and the eccentricity, and shifts the line of apsides instant by instant during one full orbital revolution. If properly accounted for in the orbital average, such features, which are of the second order in the acceleration causing them, gives rise to a further contribution of order $\mathcal{O}\ton{c^{-4}}$ to the 2PN pericentre precession which adds on the direct one. The resulting expression turns out to be dependent on the initial position $f_0$ along the orbit. Numerical integrations of the equations of motion confirmed also such a result.
Since the orbital dynamics of our Solar system is routinely modeled up to the 1PN level in harmonic coordinates of PN theory, it is unlikely that such an indirect precession can be measured separately because it does not come from a distinct acceleration which, instead, could be suitably expressed in terms of a dedicated solve-for parameter to be estimated in specific covariance analyses. For Mercury, its nominal size amounts to $16-33\,\mu\mathrm{as\,cty}^{-1}$, depending on $f_0$. For the binary pulsars, the experimental approach is different. It implies the determination, in a phenomenological, model-independent way, of several post-Keplerian parameters, among which there is also the periastron precession, from a confrontation of an analytical timing formula with the recorded pulses. Then, model-dependent analytical expressions for the measured post-Keplerian effects are used to determine the masses of the system, and to perform one or more tests of the model of gravitation considered.  In the case of PSR B1913+16, the indirect 2PN precession  ranges from $-0.000048\,\mathrm{deg\,yr}^{-1}$ to $0.001052\,\mathrm{deg\,yr}^{-1}$, while for PSR J07373039A/B it is $0.00092-0.00132\,\mathrm{deg\,yr}^{-1}$. This shows that the choice of $f_0$ may enhance or even cancel out the overall 2PN periastron precession. For OJ 287, it ranges from $20\,\mathrm{deg\,cty}^{-1}$ to $516\,\mathrm{deg\,cty}^{-1}$; the 1PN perinigricon precession amounts to  $206.8\,\mathrm{deg\,cty}^{-1}$.

We compared our formulas to some other analytical results in the literature by showing that the latter ones disagree with ours and with our numerical integrations of the equations of motion. It appears that the source of discrepancy relies in the treatment of the indirect effects arising from the inclusion of the instantaneous 1PN changes of the semimajor axis and eccentricity in the integration over one orbital revolution of the pericentre shift due to the 1PN acceleration itself.

\section*{Acknowledgements}
I am grateful to all the referees and the Editorial Board’s members who took part in the
reviewing process for their patience and constructive comments which notably improved the manuscript.

\bibliography{2PN}{}

\end{document}